# Stimuli-responsive behavior of PNiPAm microgels under interfacial confinement

Johannes Harrer [a,‡], Marcel Rey [a,‡], Simone Ciarella [b], Hartmut Löwen [c], Liesbeth M. C. Janssen [b], and Nicolas Vogel [a,*]

[a] Institute of Particle Technology, Friedrich-Alexander University Erlangen-Nürnberg, Cauerstrasse 4, 91058 Erlangen, Germany

[b] Theory of Polymers and Soft Matter, Department of Applied Physics, Eindhoven University of Technology, P. O. Box 513, 5600MB Eindhoven, The Netherlands

[c] Institut für Theoretische Physik II: Weiche Materie, Heinrich-Heine-Universität, D-40225 Düsseldorf, Germany

**Abstract**

The volume phase transition of microgels is one of the most paradigmatic examples of stimuli-responsiveness, enabling a collapse from a highly swollen microgel state into a densely coiled state by an external stimulus. Although well characterized in bulk, it remains unclear how the phase transition is affected by the presence of a confining interface. Here, we demonstrate that the temperature-induced volume phase transition of poly(N-isopropylacrylamide) microgels, conventionally considered an intrinsic molecular property of the polymer, is in fact largely suppressed when the microgel is adsorbed to an air/liquid interface. We further observe a hysteresis in core morphology and interfacial pressure between heating and cooling cycles. Our results, supported by molecular dynamics simulations, reveal that the dangling polymer chains of microgel particles, spread at the interface under the influence of surface tension, do not undergo any volume phase transition, demonstrating that the balance in free energy responsible for the volume phase transition is fundamentally altered by interfacial confinement. These results imply that important technological properties of such systems, including the temperature-induced destabilization of emulsions does not occur *via* a decrease in interfacial coverage of the microgels.

**Introduction**

Since the initial discovery that solid particles can stabilize emulsions by Ramsden and Pickering in 1903,[1,2] the behavior of particles at liquid interfaces continues to be an important multidisciplinary subject.[3–5] On the one hand, colloidal crystals at interfaces provide elegant model systems to study fundamental physical phenomena such as crystallization, phase behavior and defect structures.[6–8] On the other hand, control of particles at interfaces has given rise to a range of important technological breakthroughs such as Pickering emulsions with long term stability[2,9], the development of liquid marbles[10,11], or particle-stabilized foams.[12,13]

In contrast to hard particles, soft microgels are significantly deformed at liquid interfaces and form a distinct core-corona structure.[14,15] Their physicochemical properties are dominated more by compressibility and steric interactions between extended polymer coronas compared to their incompressible colloidal analogues and therefore show a more complex phase behavior, both at the interface[16–20] and in bulk solutions.[21–24]

Microgels are crosslinked swollen polymer networks with a diameter in the nano- or micrometer range.[25–27] A prominent backbone-crosslinker combination for microgels is poly(N-isopropylacrylamide) (PNiPAm), crosslinked with N,N'-methylenbis(acrylamide) (BIS).[28] The macromolecule thus combines hydrophilic amide and hydrophobic isopropyl groups. The competition between these groups leads to a temperature-dependent transition from an expanded coil to a globular state in aqueous environments when the temperature is raised above the lower critical solution temperature, in this case 32 °C.[29–34] At the critical temperature, the nonpolar groups aggregate and expel water from the macromolecule to increase the entropy of the system. Consequently, crosslinked microgels undergo a volume phase transition from a swollen to a collapsed state at the volume phase transition temperature (VPTT) of 32 °C.[25,28,30,35]

Typically, PNiPAm microgels are synthesized in a one-pot reaction, which produces a crosslinking gradient from the center towards the periphery of the particle and results in a relatively stiff core surrounded by loosely crosslinked dangling chains.[36–38] At the interface the core deforms and the dangling bonds spread out to form a corona, leading to a characteristic "fried-egg" morphology.[14,15,39] Small angle neutron scattering has revealed that this corona consists of a few-nanometer thin film of polymer with a low water content, which separates the cores of neighboring particles and thus prevents close packing.[40] The core elastically deforms under the effect of surface tension but remains swollen,[15,39] while the remaining dangling chains that are not adsorbed to the interface extend into the bulk liquid.[40]

The ability of microgels to conform and thus efficiently cover a liquid interface can be used to create very stable emulsions.[26,41–43] Exploiting stimuli-responsive properties of the microgels enables to break such emulsions on demand by changing the temperature or pH.[42–56] The rupture of such emulsions can be correlated with the volume phase transition of the stabilizing microgels.[42,43,49–52] Rheological studies suggest that the microgels collapse to a smaller diameter at a liquid interface, leading to a reduced interfacial coverage and thus a reduced emulsion stability.[51,52] While this model correlates experimentally observed rheological properties with the macroscopic emulsion stability, an experimental visualization to provide a microscopic picture of the microgel collapse at liquid interfaces is still missing.

Here, we correlate the interfacial behavior of PNiPAm microgels adsorbed at the air/water interface with an ex-situ, microscopic structural analysis of the microgel morphology at different temperatures to provide a detailed picture of the collapse of individual microgels at interfaces. We use a Langmuir-Blodgett technique to control the water subphase temperature, monitor the surface tension and deposit the microgel arrays from the interface to a solid substrate. These deposited microgel arrays are then characterized ex situ by scanning electron microscopy (SEM) and atomic force microscopy (AFM). We apply temperature gradients including heating, cooling, and cycling, and correlate the speed of the Langmuir-Blodgett transfer with the temperature change so that the temperature-dependent phase behavior of the microgels at the interface is directly encoded in the lateral position of the microgels on the solid substrate.[17,19] This procedure allows the observation of structural changes in the assembly of the microgel arrays as well as changes in the morphology of individual microgels. We find that the "fried-egg" or core-corona shape of the microgels adsorbed at the air/water interface persists even at temperatures much above the VPTT, independently of whether the subphase was heated or cooled, indicating that the collapse of microgels is at least partially hindered by the liquid interface. We propose a detailed model accounting for the striking difference in core and corona behavior, which we corroborate by Molecular Dynamics simulations.

**Results**

**Comparison between bulk and interface**: We first evaporate drops of a dilute microgel dispersions at different temperatures to investigate the impact of temperature on the morphology of dried PNiPAm microgels on a substrate. When dried on a heating plate at 200 °C the water evaporates and the microgels adsorb in the collapsed state from bulk to the wafer. In the dried state they have a height of 130 nm and a width of 336 nm (Figure 1, black, Suppl. Figure 1). In microgel dispersion droplets dried at room temperature, the microgels adsorb to the air/water interface and from there deposit onto the substrate after water evaporation.[57] They are more flattened and spread in the dried state (Figure 1, grey, Suppl. Figure 1) with a height of 31 nm and width of 716 nm, as expected from their typical deformed morphology at the interface.[14,15,39]

To investigate the effect of an interface on the volume phase transition of microgels, we first spread a hot microgel dispersion (80 °C) – in which the individual microgels are all in the collapsed state – onto hot water surfaces with different temperatures at a surface pressure of 5 mN/m, expecting to find collapsed microgels with similar dimensions as in the bulk phase.[51,52]. Instead, for all temperatures above the bulk VPTT, the height of microgels present at the air/water interface is significantly lower compared to microgels deposited from bulk (Figure 1). Interestingly, we observe a dependence of the height of the microgels, which increases from 54 nm at 50 °C to 72 nm at 80 °C, without reaching the bulk value. These experiments clearly indicate a different behavior compared to the collapse of microgels in bulk above the VPTT, where additional heating does not induce any further shrinking (Suppl. Figure 2).

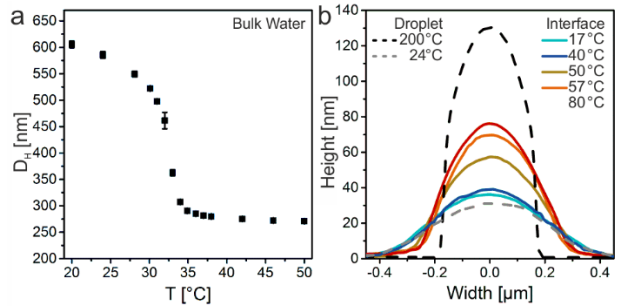

**Figure *1***: Comparison between the volume phase transition of PNiPAm microgels in bulk and at the air/water interface. a) Hydrodynamic diameter ($D_H$) – temperature diagram measured by dynamic light scattering. $D_H$ remains constant above the VPTT b) AFM cross-sections of PNiPAm microgels dried from bulk and deposited from the air/water interface at a surface pressure of 5 mN/m at different subphase temperatures. Dried from an evaporating drop at 200 °C (black) and at 24 °C (grey). Microgels deposited from the air/water interface at 17 °C (light blue), 40 °C (dark blue), 50 °C (yellow), 57 °C (orange) and 80 °C (red). The microgel morphology depends on the subphase temperature.

**Heating and cooling of microgels adsorbed at the air/water interface**: In order to investigate the morphology change of the microgels at the volume phase transition directly at the interface, we subsequently expose the adsorbed microgels to a gradual increase or decrease in temperature. To this end, we perform a slow Langmuir-Blodgett transfer to a solid substrate while continuously heating or cooling the subphase, enabling us to correlate the position of the transferred microgel layer with the temperature and the corresponding surface pressure at the interface.[17,58] These transferred microgel layers are then characterized in detail using SEM and AFM techniques. Our methodology assumes that the microgel layer is deposited onto the solid substrate without changing its conformation and structure, which may lead to a systematical error. However, we believe that this assumption is justified for the following reasons. First, the morphologies we observe in the AFM images closely agree with the typical fried-egg shape revealed by direct visualization techniques at the air/water interface.[14] Second, while the drying will certainly change the dimensions of the microgels, especially in the swollen state, we compare all the following data in the same state and against a dried sample from bulk, so that these comparisons are coherent in themselves. Third, in all cases, a crystalline hexagonal structure of the transferred monolayer is observed without any irregularities or lattices misorientations that would be the consequence of capillary forces deforming the morphology upon drying. Finally, we find a clear difference in the morphology of microgels dried from an interface compared to microgels dried from a microgel dispersion drop casted onto the substrate in hot state, indicating that the pronounced core-corona shape can indeed be attributed to their interfacial morphology (Suppl. Fig. 1g).

We first investigate the thermoresponsive interfacial behavior of the PNiPAm microgels at low surface pressure (<1 mN/m), where the surface coverage is not yet complete and the interfacial monolayer contains empty areas. The individual microgels are thus able to adapt and deform upon changes in temperature with minor interference of neighboring particles. We spread the microgels at the air/water interface at 22 °C and slowly (0.67 °C/min) heat the water subphase while simultaneously performing the Langmuir-Blodgett transfer. If the collapse occured similar to the bulk phase with the microgels changing from the fried-egg shape into a dense sphere, we would expect a decrease in the particle diameter visible in the SEM images combined with an increase in microgel height measured by AFM. The magnitude of this transformation would be comparable to the change seen in Figure 1. Furthermore, the decrease in microgel size would lead to a melting of the crystalline hexagonal arrangement at the interface, similar to previous studies on crystal phase transitions observed in bulk.[59–61]

The data, presented in Figure 2a, reveals a strikingly different behavior. The interfacial microgels show only a change in core diameter by 7 % with a decrease of 50 nm upon heating (Figure 2a, red data points). These values correspond to a change in cross-sectional area by a factor 1.15, which much less pronounced compared to the bulk, where the cross-sectional area changes by a factor 5 (Suppl. Figure 2). AFM measurements, showing only minor changes in particle height at different temperatures, confirm this behavior (Figure 2b). Additionally, the interparticle distance does not change with increasing temperature (Figure 2c). The AFM phase images (Figure 2e,f) reveal that neither core nor corona collapse and both remain spread out at the interface even at 57 °C, similar to their initial state at 22 °C. Overall, under these conditions, the microgels do not collapse at the interface as expected from bulk measurements.

To fully explore the effect of a temperature change across the VPTT at the interface, we also perform the reverse experiment by first spreading a hot microgel dispersion (80 °C) onto a hot air/water interface (57 °C) and subsequently letting it cool to room temperature (Figure 2, blue data points). Remarkably, in this case, the initial diameter at 57 °C is 541 nm, which is much smaller than the diameter of the particles that are spread cold and heated to the same temperature (Figure 2a). The height of the microgels is 62 nm, which is 25 nm higher compared to the reverse heating experiment (Figure 2b). Both height and diameter, however, still do not reach the values of the collapsed particles deposited from the bulk. During cooling, the core diameter increases to the same value as for the microgels added onto the cold interface when the temperature decreases below 32 °C (Figure 2a). The microgel height reflects this behavior and decreases from 62 nm at 57 °C to 32 nm at 22 °C, which is the same value as for the microgels spread to the cold interface. We further extracted the dry volume of the microgels from AFM images which remained constant at 5.5 * $10^{-21}$ m$^3$. An increase in the interparticle distance is observed below 32 °C (Figure 2c), clearly showing that under these conditions, the microgels expand at the VPTT of 32 °C. In the AFM phase images, however, a clear core-corona structure persists throughout the experiment. Importantly, while the collapsed core region expands upon cooling, the corona of expanded chains is already present at high temperatures and does not change upon temperature decrease (Figure 2g,h). We extract the core diameter and the diameter of the core plus corona from the AFM height and phase images, with which we calculate the corona area throughout the experiment (Figure 2d). The corona area remains constant and does not significantly change through the temperature range. Hence, we conclude that the increase in interparticle distance is solely caused by the expansion of the core.

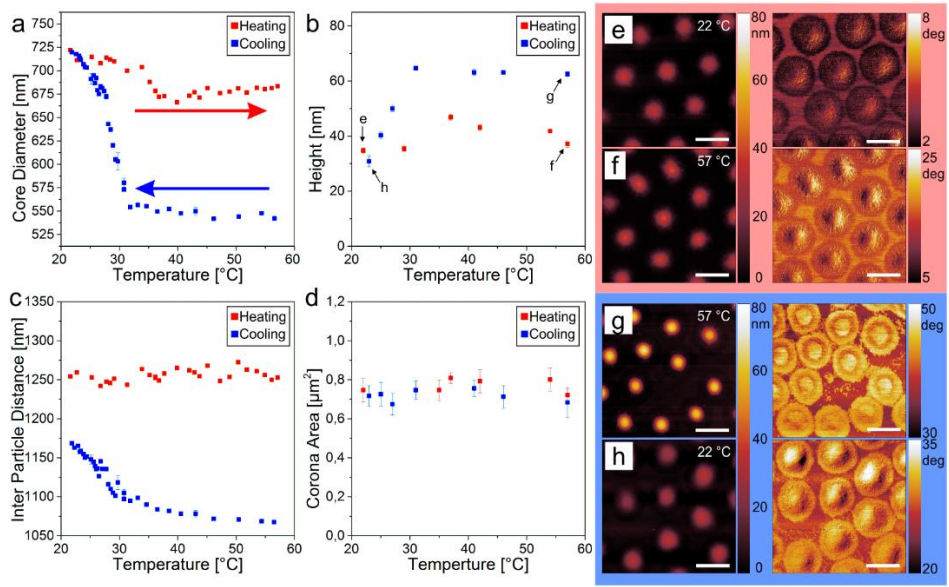

**Figure 2**: Temperature-dependent swelling behavior of PNiPAm microgels adsorbed at the air/water interface for heating and cooling at a surface pressure of 0.5 mN/m a) Core diameter vs temperature diagram. A hysteresis between heating and cooling can be observed. b) Average height measured with the AFM in dependence of the subphase temperature. c) Temperature-dependent changes in interparticle distance. The distance increases while cooling but is unaltered while heating the microgels d) Corona area vs temperature, extracted from AFM phase images. In both cases, the corona area remains unaffected by temperature changes. e,f) Corresponding AFM height and phase images at 22 °C and 57 °C when the particles are heated. g,h) Corresponding AFM height and phase images at 57 °C and 22 °C when the particles are cooled. Scale bar: 1 μm

Next, we repeat the heating and cooling experiments for a high surface pressure (~16 mN/m), where the monolayer is fully closed and collective effects of the volume phase transition on the surface pressure can be recorded. During heating of microgels spread on a cold air/water interface, the core diameter decreases by 50 nm (Figure 3a), similar as above, while the height of the microgels increases by 19 nm (Figure 3b). Noteworthily, the height profiles in Figure 3b are generally 10 nm higher than for low surface pressures (Figure 2b), as the higher packing density causes a compression of the microgels.[62] The interparticle distance remains constant at all temperatures (Figure 3c) and we do not observe any melting or reordering of the crystal lattice. This implies that the corona is unaffected by the temperature increase, thus keeping the individual cores at a uniform distance.

Reversing the temperature ramp by applying the microgels to a hot interface and subsequently letting the subphase cool to room temperature again reveals notable differences compared to the heating experiments. Both the initial core diameter and the height of the microgels at the hot interface are markedly distinct between particles that are first spread at room temperature and subsequently heated to the same final temperature (Figure 3a,b): the initial core diameter is now significantly lower (d = 500 nm), while the microgel height at 57 °C is significantly higher (h = 80 nm). After cooling, both the diameter and height become similar to the values found for microgels spread at room temperature (d = 625 nm and h = 42 nm, respectively). In all cases, the interparticle distance does not change with temperature (Figure 3c). The surface pressure-temperature diagram (Figure 3d) displays a hysteresis with a minimum at 32 °C during cooling, while remaining constant above 32 °C during heating. In summary, we observe a hysteresis in all relevant parameters between heating and cooling cycles.

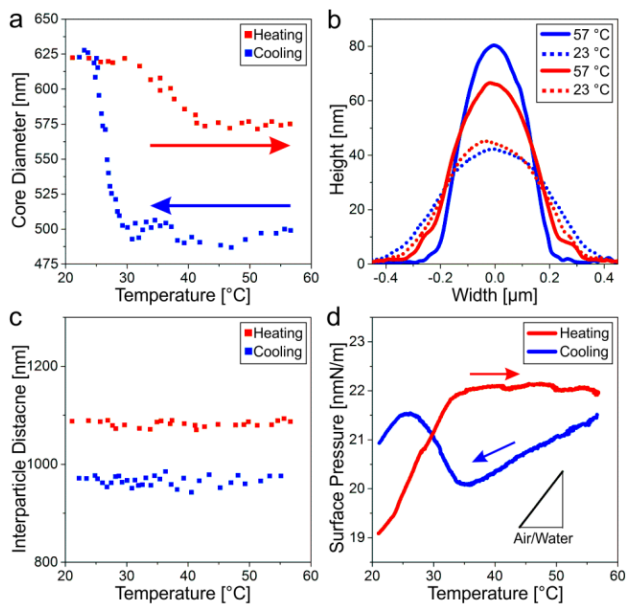

**Figure 3**: Temperature-dependent swelling behavior of PNiPAm microgels adsorbed at the air/water interface for heating (red curves) and cooling (blue curves). a) Core-diameter vs temperature. A hysteresis

between heating and cooling can be observed. b) AFM height cross-sections. c) Temperature-dependent change in interparticle distance. d) Change in surface pressure when the microgels are heated and cooled at the interface. The inset shows the slope of surface pressure change with increasing temperature of a pure air/water interface.

**Temperature Cycling**: Let us now investigate the observed hysteresis in more detail *via* temperature cycling experiments. First, we spread the microgels at 57 °C, cool the subphase to 21 °C, and subsequently perform multiple heating-cooling cycles. Figure 4a shows the correlation between temperature and core diameter over time during the cycling experiments. When the particles are initially spread to the hot interface, their core diameter is the smallest. Upon cooling, this core diameter expands, but never collapses to the initial value in the subsequent heating cycles. Instead, the observed difference in core diameter becomes smaller with each heating cycle. The height of the particles during the temperature cycling (Suppl. Figure 3) shows a similar behavior. Importantly, throughout the entire cycling experiment, we do not observe any changes in interparticle distance (Figure 4b) or a rearrangement in the monolayer structure. This behavior implies that the expanded coronae remain in contact and effectively jam the interface, thus preventing any structural reorganization even as the microgel cores change their dimensions.

The evolution of surface pressure throughout the cycling experiment (Figure 4c) shows a pronounced hysteresis between the first cooling and re-heating, while subsequent temperature cycles are reversible and do not recover the initial cooling behavior. As detailed below, this behavior indicates that the microgels spread in the collapsed state to the hot interface are kinetically trapped in an out-of-equilibrium state, which cannot be recovered after relaxation at low temperature and reheating.

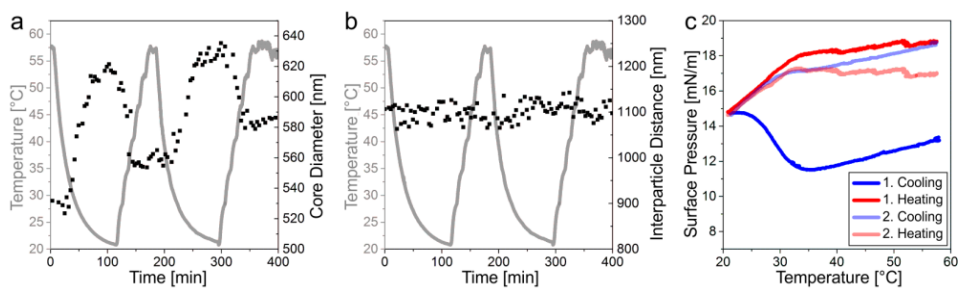

**Figure 4**: Temperature cycling of microgels adsorbed at the air/water interface spread at 57 °C. a) Development of temperature (grey lines) and core diameter (black points) over time. b) Development of temperature (grey lines) and interparticle distance (black points) over time. c) Surface pressure – temperature diagram for two consecutive cooling and heating cycles. A rise in surface pressure can be seen below 32 °C for the first cooling cycle. Further heating and cooling cycles show the same surface pressure development as in the 2$^{nd}$ cycle.

**Molecular Dynamics simulations**: To further investigate the volume phase transition at the interface, we employ Molecular Dynamics simulations to synthesize and model the microgel particles *in silico*. The microgel architecture is represented by Finite-Extensible-Nonlinear-Elastic (FENE) bonds combined with Weeks-Chandler-Andersen (WCA) crosslinking interactions.[63,64] We choose a crosslinker-to-monomer ratio of 4 %, close to the experimentally studied PNiPAm morphology. The thermoresponsivity of the microgel particles is achieved through an additional potential that mimics the solvent.[65] We first reproduce the bulk volume phase transition by changing the effective temperature parameter $\alpha$,[38] such that the net interactions between the individual microgel monomers switch from repulsive (swollen state) to attractive (collapsed state)[63] (Figure 5a,b(i)), leading to a similar volume phase transition as measured experimentally in Suppl. Figure 2a. Next, we let the microgels in the swollen state adsorb to the air/water interface by introducing an external Lennard-Jones wall potential along the Cartesian z-direction, centered on the z = 0 plane and oriented such that negative *z* corresponds to the repulsive air phase and positive *z* to the bulk water phase (Suppl. Figure 4.). The attractive minimum of the external potential corresponds to the air/water interface, thus mimicking the surface tension pulling on the particles (Figure 5a(ii)). We find that the microgel core at the interface remains swollen and deforms laterally. The dangling chains near the air/water interface stretch out to form the corona while the remaining dangling chains extend into the bulk water phase, reproducing the typical "fried-egg" morphology known from experiments.[14] We then increase the effective temperature to above the VPTT (Figure 5a(iii)). The adsorbed microgel core as well as the dangling chains extending into the water subphase exhibit a collapse, but the microgel remains notably deformed at the interface. Importantly, the dangling chains in the corona region do not change their state and remained stretched at the interface. This behavior reproduces our experimental finding that the corona persists even above the VPTT. Finally, we also perform the reverse numerical experiment by letting bulk-equilibrated collapsed microgels (Figure 5b(i)) adsorb to the interface through the addition of the external potential (Figure 5b(ii)). The chains in the microgel core region remain collapsed, but are nevertheless significantly deformed under the effect of the potential mimicking the surface tension. Additionally, a clear corona region of dangling chains stretched out along the interface forms upon adsorption, even though these chains were initially collapsed into a globular structure in the bulk. Upon reducing the effective temperature of the microgel at the interface, the core swells and expands and dangling chains extend into the water subphase (Figure 5b(iii)). Together, these Molecular Dynamics simulations of the microgel model reproduce the essential findings of the experiments.

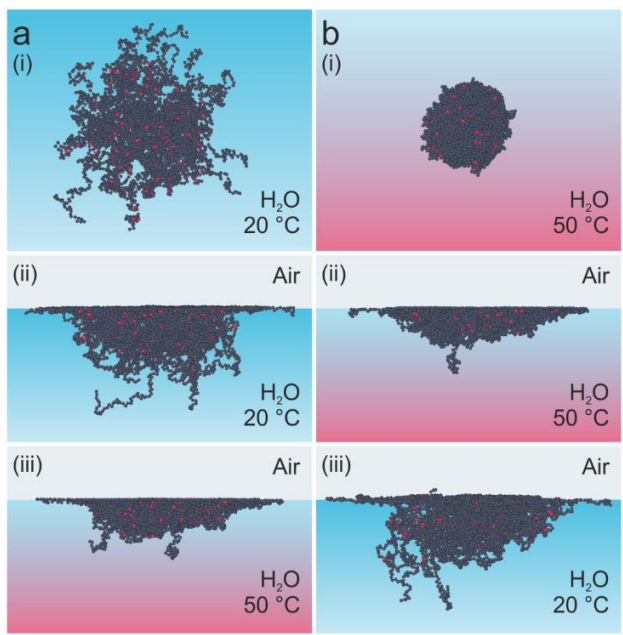

**Figure 5**: Molecular Dynamics simulations of microgels synthesized *in silico* in bulk and adsorbed to the air/water interface before and after the volume phase transition, modeled using a combination of FENE, WCA, thermoresponsive, and interfacial potentials. Red beads correspond to crosslinkers. The phase transition in the model occurs at an effective temperature of $\alpha$, which corresponds to ~35 °C in the scale of the figure. a(i) The microgel forms a spherical, swollen state in the bulk, mimicking the morphology at 20 °C. The phase transition to a collapsed state is induced by increasing the effective temperature $\alpha$ to 1, thus mimicking an attractive behavior of the monomers (b(i)). Introduction of an artificial interface, described by a one-dimensional Lennard-Jones potential to which the microgel is attracted, causes a deformation of the microgel and the formation of a core-corona morphology (a(ii)). Inducing the volume phase transition of the microgel at the interface (a(iii)) by changing the monomer interaction similar to the bulk case leads to a collapse of the core and dangling chains that were initially extended into the subphase, yet the corona persists. b(ii)) Introducing the artificial interface to adsorb the microgel in the collapsed state leads to a similar morphology with a deformed core and an extended corona. b(iii)) Subsequent transitioning into a swollen state at the interface shows an expansion of the core region and dangling chains extending into the subphase.

**Discussion**

The microscopic investigation of the morphology of PNiPAm microgels at the interface at different temperatures, albeit assessed indirectly via a transfer to a solid substrate, indicates that the volume phase transition is strongly affected by the interface. In contrast to the bulk behavior, microgels at the air/water interface show a hysteresis between heating and cooling, and an incomplete core collapse – leading to different dimensions at different temperatures, and, most importantly, a corona that remains extended at all experimentally assessed temperatures. These observations contrast existing models and necessitate a detailed, molecular interpretation of the role of surface tension on the different parts of the microgel. As we will see below, our interpretation naturally explains the established macroscopic properties of stimuli-responsive microgels such as the destabilization of emulsions upon heating.

We first briefly recapitulate that our synthesized PNiPAm microgels consist of a relatively densely crosslinked core and a more loosely crosslinked shell with dangling polymeric chains at the outside[36,38] and that they deform into a core-corona structure at the interface.[14,15,39] At low surface pressures, such microgels assemble into a hexagonal non-close-packed lattice, where they are in corona-corona contact.[17] As corroborated by a recent neutron reflectivity study, PNiPAm microgels adsorbed at the air/water interface can be divided into three structural regimes[40]: (i) The corona, consisting of a nanometer thin film of highly stretched PNiPAm chains with a low water content; (ii) the more crosslinked, fully solvated microgel core; and (iii) dangling polymer chains extending into the bulk solution (Figure 6a). A similar model was also proposed recently for PNiPAm microgels adsorbed to a silicon wafer investigated by grazing incidence small-angle neutron scattering.[66] In the following, we discuss the effect of temperature on these three regimes at the interface individually.

**The corona structure persists above the VPTT:** In contrast to SEM or AFM height images that cannot detect the corona of a microgel because of its extremely small height,[40] the phase contrast imaging in AFM is sensitive enough to visualize the corona.[67] In all phase images, we detect an extended corona surrounding the microgel core at the interface, regardless of the temperature or preparation of the system (Figure 2e-h).

Furthermore, the area of the corona remains similarly unaffected by changes in temperature and thus prevents a change in interparticle distance in the interfacial layer of microgels upon heating or temperature cycling (Figure 3c, Figure 4b). The persistence of the corona at all temperatures regardless of the system preparation is entirely reproduced by our computer simulations (Figure 5), similar to previous simulations on polymer-grafted nanoparticles.[68] We only observe a change in the microgel lattice in the first cooling cycle of microgels spread at a hot interface. We attribute this behavior to an expansion of the microgel core and discuss the effect in the context of the hysteresis between heating and cooling below. We rationalize the unexpected behavior of the corona in terms of the conformation of dangling polymer chains at the interface. These chains have a much lower water content compared to dangling chains in the bulk subphase, because the hydrophobic groups of the PNiPAm polymer orient themselves towards the air phase to reduce the surface tension. The hydrophobic groups are therefore not surrounded by oriented water molecules even below the VPTT. Additionally, the conformation of the polymer chains at the interface within the corona is more stretched compared to its bulk conformation, as the decrease in surface tension upon elongation of the polymer chains competes with the tendency to form a Gaussian coil conformation to maximize entropy. This stretched nature of the corona and the low water content has been confirmed by neutron reflectivity experiments.[40] Together, these effects lower the gain in free energy upon chain collapse, which is driven by an increase in entropy upon volume reduction caused by the release of oriented water around the hydrophobic groups. In summary, we attribute the absence of any detectable temperature-induced collapse of the corona region to the additional contribution of surface tension to the free energy of the microgels (Figure 6b).

**Temperature-induced change in microgel core size:** We observe a change in core diameter and height of the microgels at the interface as the temperature changes. However, this change in dimensions throughout all experiments is significantly lower than expected from the bulk behavior (Figures 1,2,3). We also attribute this decreased volume phase transition to the influence of surface tension at the interface. The microgel chains in the core are in the swollen state below the VPTT and can reduce their free energy by releasing oriented water molecules around the hydrophobic moieties.[30,33] However, surface tension adds an additional term to the free energy which drives the microgel core to deform and increase its area at the interface, as has been shown both in experiments and simulations,[14,15,39,69] and is reproduced in our simulations (Figure 5). Thus, surface tension at the air/water interface partially counteracts the tendency to collapse in the lateral direction; indeed, our results strongly suggest that microgels adsorbed to the interface predominantly shrink only in the $z$-direction, making the effective change in microgel core diameter and height across the VPTT much smaller than in bulk (Figure 6b).

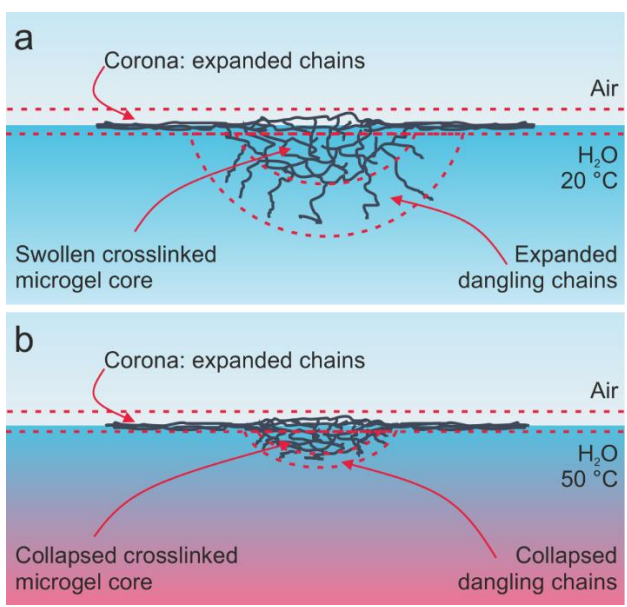

**Figure 6**: Schematic illustration of PNiPAm microgels adsorbed to the air/water. a) Below the VPTT, they exhibit a corona from expanded, dangling chains at the air/water interface, a swollen crosslinked core, and dangling chains extending into the water phase. b) Above the VPTT, we expect both core and dangling chains to be collapsed while the corona remains unaffected.

**Hysteresis of the volume phase transition:** Throughout our experiments, we observe a clear difference in behavior between microgels that are initially spread onto a cool subphase and subsequently heated, and microgels that are spread onto a hot subphase and exposed to the reverse temperature gradient (Figures 2-4). From the temperature cycling experiments shown in Figure 4, we conclude that the first cooling cycle shows an apparent irreversible behavior, while all subsequent cycles are manifestly reversible. This first cooling of microgels spread at a hot temperature leads to a different surface pressure – temperature behavior (Figure 3,4) and shows a larger height of the microgels compared to the consecutive heating and cooling cycles (Figure 3,4), marking a clear hysteresis.

We attribute this hysteresis between cooling and heating cycles to kinetic trapping of the microgels spread on a hot subphase, as explained below. First consider that a microgel spread in the collapsed state will resist deformation caused by the surface tension more effectively than a swollen microgel, because of the aggregation of the hydrophobic groups within the polymer that causes a higher elastic modulus in the collapsed state.[15,39,70,71] Since the air/water surface tension decreases with temperature, so does the strength of its contribution to the free energy. Therefore, it is expected that the change in dimensions (increase in height) upon collapse will be stronger with increasing temperature, which is indeed found for microgels spread at different temperatures above the VPTT as shown in Figure 1, and agrees with literature reports on the microgel morphology upon spin- and dip coating onto solid substrates.[72]

Upon decreasing the temperature below the VPTT, swelling of the microgel core will decrease the Young's modulus and thus increase the deformability, leading to the observed increase in core diameter, and decrease in surface tension (Figure 3). Additionally, provided there is sufficient space at the interface to expand (low surface pressure, Figure 2), the interparticle distance also increases. However, upon heating from a cold subphase, the swollen microgel is already expanded at the interface. The collapse is therefore confined to neighboring hydrophobic groups and prevents the microgels from reaching the more compact conformation observed after hot-spreading, which causes the observed hysteresis.

**Influence of temperature on the interfacial tension:** PNiPAm microgels are known to lower the surface tension of water.[73] The effect of temperature on the interfacial tensions was investigated for air/water[29,51,74] and oil/water[52,56] interfaces, either by the pendant drop method[29,52,56,74] or on a Langmuir trough.[51] The experiments revealed a minimum in the surface tension around the VPTT of the microgels[29,51,52,56,74] as well as a hysteresis between heating and cooling of an individual droplet,[52,56] similar to the results of our experiments (Figure 4c). In order to compare the behavior of microgels at oil/water and air/water interfaces we repeated the temperature gradient experiment for oil/water interfaces (Figure 7). Microgels adsorbed at the oil/water interface (Figure 7) show a qualitatively similar behavior in interfacial tension and morphology compared to microgels adsorbed at the air/water interface (Figure 3). In earlier work, the minimum in surface tension around the VPTT was assigned to a change in interfacial coverage around the VPTT[51,52] or to aggregation of the microgels at the VPTT.[29,56] In contrast to the pendant drop experiment, where the microgel adsorption kinetics may change with temperature and thus the number of microgels at the interface may change dynamically,[29,52,56,74] our system keeps this number fixed as the microgels are spread at the interface and are not expected to desorb.[75] The constant interparticle distance during the cycling experiment (Figure 3, Figure 4 and Figure 7) supports this claim. Furthermore, the microgel corona at the interface does not collapse at different temperatures and remains spread at the interface, thus covering an essentially constant surface area both above and below the VPTT (Figure 2). We may therefore conclude that the interfacial coverage in our experiments remains constant.

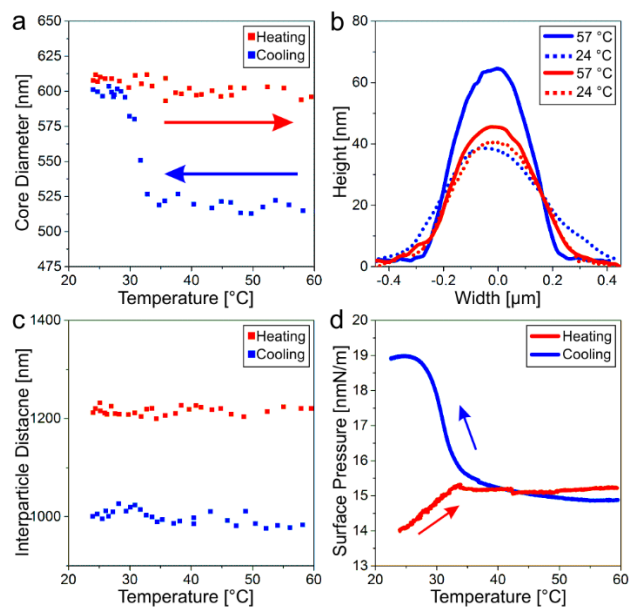

**Figure 7**: Temperature-dependent swelling behavior of PNiPAm microgels adsorbed at the dodecane/water interface for heating (red curves) and cooling (blue curves). a) Core-diameter vs temperature. A hysteresis between heating and cooling can be observed. b) AFM height cross-sections. c) Temperature-dependent

change in interparticle distance. d) Change in surface pressure when the microgels are heated and cooled at the interface. We observe a qualitatively similar behavior compared to microgels adsorbed at the air/water interface (Figure 3).

To investigate whether the microgel architecture (core or corona) has an effect on the interfacial behavior, we repeated the cycling experiment using linear polyNiPAm chains (40 kD), which serve as a mimic of the non-crosslinked, dangling chains of the corona. This experiment is therefore expected to isolate the interfacial properties of the corona from that of the crosslinked and deformed microgel core. Linear PNiPAm chains closely reproduced the surface pressure – temperature behavior of the microgels, including the hysteresis between heating and cooling (Figure 8). These results imply that the surface pressure evolution as a function of temperature does not depend on the microgel architecture but is rather an inherent characteristic of the PNiPAm polymer. The change in surface tension with temperature, becomes different when water-soluble surfactants are present.[76] This slope in the surface tension-temperature diagram is characteristic for individual surfactants adsorbed to the air/water interface.[76] While a detailed molecular interpretation of the temperature-dependent interface behavior of linear PNiPAm chains is not within the scope of this article, the temperature-dependent surface pressure may be qualitatively interpreted as a change in the surfactant properties of the PNiPAm due to the change in molecular structure upon collapse.

Our results suggest that the hysteresis and the temperature-dependence of the surface pressure is governed by the characteristic behavior of linear poly(NiPAm) chains rather than the microgel architecture.

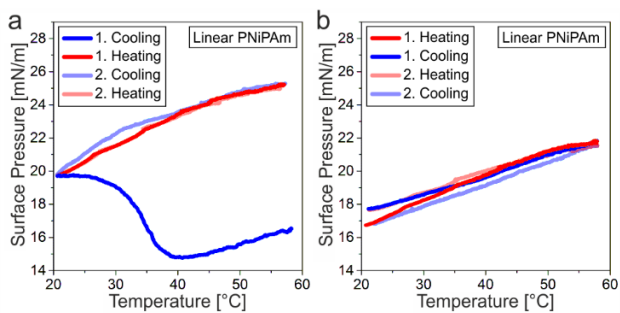

**Figure 8**: Surface pressure – temperature diagram of linear PNiPAm chains (40 kD) at the air/water interface during temperature cycles spread at 58 °C (a) and 21 °C (b).

**Impact on stimuli-responsive emulsions:** Finally, we discuss how our findings can be interpreted in the context of the destabilization of stimuli-responsive emulsion. PNiPAm microgels are known to stabilize oil/water emulsions, which are stable below the VPTT and can be destabilized by heating above the VPTT.[42,43,52–56,44–51] Various explanations regarding this stimuli-responsive behavior were proposed. One model attributes the destabilization of the emulsion to the collapse of the interfacially adsorbed microgels into smaller spheres which cover less area and thus are less efficient stabilizers.[51–53] Additionally, it was proposed that microgels may desorb from the oil/water interface to the oil phase due to their increased hydrophobicity,[53,54] which, however, does not seem to occur in all reported experimental scenarios.[55,56] It was also suggested that microgels form aggregates and multilayers at the oil/water interface,[44,45,56] which has been experimentally visualized for the inverse case of water in oil emulsions.[44] Last, the destabilization of the emulsion was attributed to a change in viscoelastic properties of the PNiPAm microgels adsorbed at the interface above the VPTT.[45,54,55,77,78]

These different explanations demonstrate the complexity of the stimuli-responsive emulsion destabilization. Furthermore, the mentioned studies on stimuli-responsive emulsion typically used around 1 wt-% microgels in the aqueous phase, which, depending on the swelling ratio, can correspond to up to 20 vol-%. It is thus likely that interactions of bulk microgels with the interfacially-adsorbed species, for example the formation of aggregates or multilayers, may affect the stability as well.[44,45,56] In addition, subsequent microgel adsorption upon temperature change is possible.[44,56]

In our work, we investigated an "ideal" Pickering emulsion where only a monolayer of microgels is present at the interface without the possibility of further microgel adsorption or interaction with microgels dispersed in the aqueous subphase. In this highly controlled experimental scenario, we conclude that there is no decrease in interfacial coverage by either shrinking or desorption of the microgels, as previously suggested.[51–54] Our results indicate that the collapse of the microgels is hindered at the interface and the corona remains unaffected by the change in temperature, thus preventing a change in the arrangement or surface coverage. The absence of any desorption from the liquid interface is supported by the constant interparticle distance and similar morphologies of the interfacial layer throughout the temperature change (Figure 3,4,7). We point out that in the simulations, we observe a clear change in the morphology of the dangling chains extending into the subphase upon temperature change (Figure 5). These dangling chains may provide steric stabilization below the VPTT, which might be reduced upon collapse onto the microgel core above the VPTT.

**Conclusions**

In summary, we have investigated the effect of temperature on the interfacial behavior of PNiPAm microgels adsorbed at the air/water interface and compared the volume phase transition behavior to bulk microgel suspensions.

We find that the typical "fried-egg" or core-corona shape of the microgels adsorbed to the air/water interface persists even above the VPTT, independently on the subphase temperature. The presence of a corona even above the VPTT preserves the hexagonal arrangement of the microgels and prevents structural rearrangements. Furthermore, the core-corona shape hinders the collapse of the microgel core at least partially, leading to smaller changes in microgel height and diameter compared to microgels deposited from bulk. We reproduce these experimental findings by coarse-grained Molecular Dynamics simulations, revealing a core-corona morphology that is preserved upon changes in temperature.

We further find that the hysteresis in surface pressure between heating and cooling is independent of the microgel architecture, confirmed by the fact that both crosslinked and linear polyNiPAm chains show similar hysteresis patterns. We attribute this hysteresis to kinetic trapping during the spreading above the VPTT. Once equilibrated below the VPTT, the effect of heating and cooling is reversible. Our results demonstrate how the presence of an interface strongly affects the thermodynamics of such stimuli-responsive microgel particles, which may also affect more applied aspects of such systems, such as the stability and temperature-induced destabilization of emulsions.

## Experimental

*Microgel synthesis:* All microgels were synthesized by precipitation polymerization following literature protocols.[19,79] A detailed description can be found in the Supplementary Information.

*Interfacial characterization.* We used a Langmuir trough (KSV Nima) setup (Suppl. Figure 5) filled with ultrapure water with a thermostat to control the water temperature and a custom-built water level equilibration device (details in Suppl. Inform.), while the surface pressure, set to zero at 20 °C, is recorded by a Wilhelmy plate. Clean poly(N-isopropylacrylamide) microgel dispersions (details in Suppl. Inform.) were diluted to 0.05 wt% and spread with a 1:1 ratio of ethanol onto the air/water interface either in the cold, expanded state (22 °C) or in warm, collapsed state (80 °C) onto subphases with different temperatures. Next, the trough water temperature was increased/decreased (0.67 °C/min) while simultaneously depositing the interfacial arrangement onto a tilted silicon wafer (10 x 0.5 cm$^2$), which was lifted through the air/water interface by 0.2 mm/min. This allows correlating each position of the wafer to their corresponding temperature and deposition time.[17,19] We characterized the dimensions of the microgels ex-situ using scanning electron microscopy (SEM, Zeiss Gemini 500) and atomic force microscopy (AFM, JPK Nano Wizard, cantilever: Anfatec NSC 18). The AFM images were postprocessed and analyzed using Gwyddion, while the SEM images were analyzed using custom-written image analysis software (details in Suppl. Inform.).

*Computer Simulations*: Each simulated microgel particle was self-assembled *in silico* from 5750 monomers and 250 crosslinkers, which is significantly smaller than the experimental system. We intend these simulations as a coarse-grained representation of the polymer network to qualitatively support the essential findings of the experiments. The difference in system size prevents a quantitative comparison. Both monomer and crosslinker species were represented as repulsive WCA beads with respectively two or four patches that can form 1-to-1 attractive bonds. After initial equilibration, we converted all patch-patch bonds into FENE potentials, freezing the topology of our microgel particle[63] (details in Suppl. Inform.). The thermoresponsivity was implemented through an additional potential[65] that mimics the solvent. Changing its effective temperature allows us to switch from the swollen to collapsed state. The air/water interface was modeled as an external one-dimensional Lennard-Jones wall potential, with the repulsive short range corresponding to the air subphase and the minimum lying at $z = 0$. We used a depth of the energy well of 1.5 kT and a width of the attractive minimum in the order of one monomer unit. These values showed the closest agreement with the experimentally observed fried-egg morphology. All Molecular Dynamics simulations were performed in the canonical ensemble (details in Suppl. Inform.).

**ASSOCIATED CONTENT**

**Supporting Information**

The Supporting Information is available free of charge on the ACS Publications website at DOI:

Experimental details and characterization of the microgels, simulation details and experimental set-up.

**AUTHOR INFORMATION**

**Corresponding Author**

*Email: nicolas.vogel@fau.de

**Notes**

Competing interests: The authors declare no competing interests.

**ACKNOWLEDGMENT**

NV and HL acknowledge funding from the Deutsche Forschungsgemeinschaft (DFG) under grant number VO 1824/8-1 and LO 418/22-1, respectively. N.V. also acknowledges support by the Interdisciplinary Center for Functional Particle Systems (FPS).

TOC:

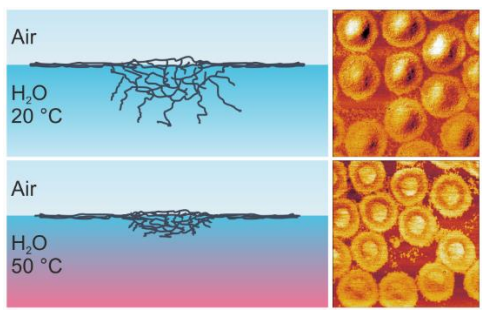

For Table of Contents use only